\documentclass[]{aa} % for the letters 
\usepackage{graphicx}
\usepackage{txfonts}
\usepackage{natbib}
\usepackage{multirow}
\usepackage{longtable}
\usepackage{pdflscape}
\usepackage{fixltx2e}
\usepackage{subfig}
\usepackage{textgreek}
\usepackage{mathtools}
\usepackage{color}
\usepackage{lmodern}
\usepackage{epstopdf}

\bibpunct{(}{)}{;}{a}{}{,}
\usepackage[colorinlistoftodos,color=blue!20]{todonotes}

\newcommand{\FIG}[1]{}

\def\mso{\,{\rm M}_\odot}

\def\lso{\,{\rm L}_\odot}

\def\kms{\, {\rm km}\, {\rm s}^{-1}}

\def\msoy{\, \mso~{\rm yr}^{-1}}

\begin{document}
   \title{ALMA detection of a tentative nearly edge-on rotating disk around the nearby AGB star R~Doradus}
   \titlerunning{ALMA detects a tentative nearly edge-on rotating disk around the nearby AGB star R~Dor}

   \author{Ward Homan
          \inst{1}
          \and
          Taissa Danilovich
          \inst{1}
          \and
          Leen Decin
          \inst{1}
          \and
          Alex de Koter
          \inst{1,2}
          \and
          Joseph Nuth
          \inst{3}
          \and
          Marie Van de Sande
          \inst{1}
          }

   \offprints{W. Homan}          
          
   \institute{$^{\rm 1}\ $Institute of Astronomy, KU Leuven, Celestijnenlaan 200D B2401, 3001 Leuven, Belgium \\
             $^{\rm 2}\ $Sterrenkundig Instituut `Anton Pannekoek', Science Park 904, 1098 XH Amsterdam, The Netherlands\\
             $^{\rm 3}\ $NASA Goddard Space Flight Center, Code 690, Greenbelt MD 20771 USA \\ 
             }

   \date{Received <date> / Accepted <date>}
 
   \abstract  
   {A spectral scan of the circumstellar environment of the asymptotic giant branch (AGB) star R~Doradus was taken with ALMA in cycle 2 at frequencies between 335 and 362 GHz and with a spatial  resolution of $\sim$150 milliarcseconds. Many molecular lines show a spatial offset between the blue and red shifted emission in the innermost regions of the wind. The position-velocity diagrams of this feature, in combination with previous SPHERE data and theoretical work point towards the presence of a compact differentially rotating disk, orientated nearly edge-on. We model the $^{\rm 28}$SiO ($v=1,~J=8\to7$) emission with a disk model. We estimate the disk mass and angular momentum to be $3 \times 10^{-6} \mso$ and $5 \times 10^{40}\ {\rm m^2 kg/s}$. The latter presents an `angular momentum problem' that may be solved by assuming that the disk is the result of wind-companion interactions with a companion of at least 2.5 earth masses, located at 6 AU, the tentatively determined location of the disk's inner rim. An isolated clump of emission is detected to the south-east with a velocity that is high compared to the previously determined terminal velocity of the wind. Its position and mean velocity suggest that it may be associated with a companion planet, located at the disk's inner rim.}
   

   \maketitle
%
%________________________________________________________________

\section{Introduction}

In their later evolutionary stages, low- to intermediate-mass stars ($\sim$0.8 - 8~$\mso$) ascend the asymptotic giant branch (AGB), where pulsational induced surface instabilities inflate their outer envelope. This envelope constitutes a dense molecular gas that partly condenses into dust grains. The solid particles efficiently absorb the incident stellar flux and are accelerated outwards, driving a stellar wind. Such AGB star outflows have been the subject of intense scrutiny, with studies aiming to for example derive mass-loss rates and abundances of molecular species \citep[e.g.][]{DeBeck2010,Decin2010}. Spherical symmetry has often been assumed when deriving the properties of this circumstellar matter. However, in this era of high spatial resolution, interferometric observations of AGB star circumstellar envelopes (CSEs) have revealed various non-spherical morphologies. These include spirals, such as those seen around the carbon stars R Sculptoris \citep{Maercker2012}, CW~Leo \citep{Decin2015} and CIT6 \citep{Kim2015}, a differentially rotating disk around the oxygen-rich AGB star L$_{\rm 2}$ Puppis \citep{Kervella2016,Homan2017}, a torus around the S-type star RS~Cnc \citep{Libert2010}, and seemingly void regions in the circumstellar medium of Mira \citep{Ramstedt2014}. 

The steadily increasing sample of AGB stars whose winds exhibit curious features, combined with theoretical support from hydrodynamical models is pointing towards binarity as one of the major wind-shaping mechanisms. The presence of a companion in the inner wind of the AGB star gravitationally affects both the medium it traverses and possibly even the behaviour of the mass-losing star, resulting in complex interactions that may manifest as spirals, clumps or disks. That the companion fraction is likely to be high is supported by the multiplicity frequency of the main sequence (MS) progenitors of AGB stars. This is found to exceed 50 percent \citep{Raghavan2010,Duchene2013}, not counting planetary companions. Though a handful of studies show that binarity can indeed conceptually explain the abovementioned features, the intricacies of these complex interactions remain unknown. Detailed hydrodynamical simulations are computationally intensive, permitting the investigation of only a few highly specific physical set-ups,  confining any detailed understanding of wind-binary interactions to these precise cases. 

Revealing and understanding the local conditions of morphologically complex winds sheds light on the processes that dictate its current steady state, as well as the further evolution into its potentially aspherical progeny. Furthermore, understanding the role planetary companions play in this picture will yield new constraints on the morphological and late phases of evolution of stars and their planetary systems.

\begin{figure*}[]
% \centering
% \floatbox[{\capbeside\thisfloatsetup{capbesideposition={left,top},capbesidewidth=4cm}}]{figure}[\FBwidth]
\sidecaption
\includegraphics[width=0.70\textwidth]{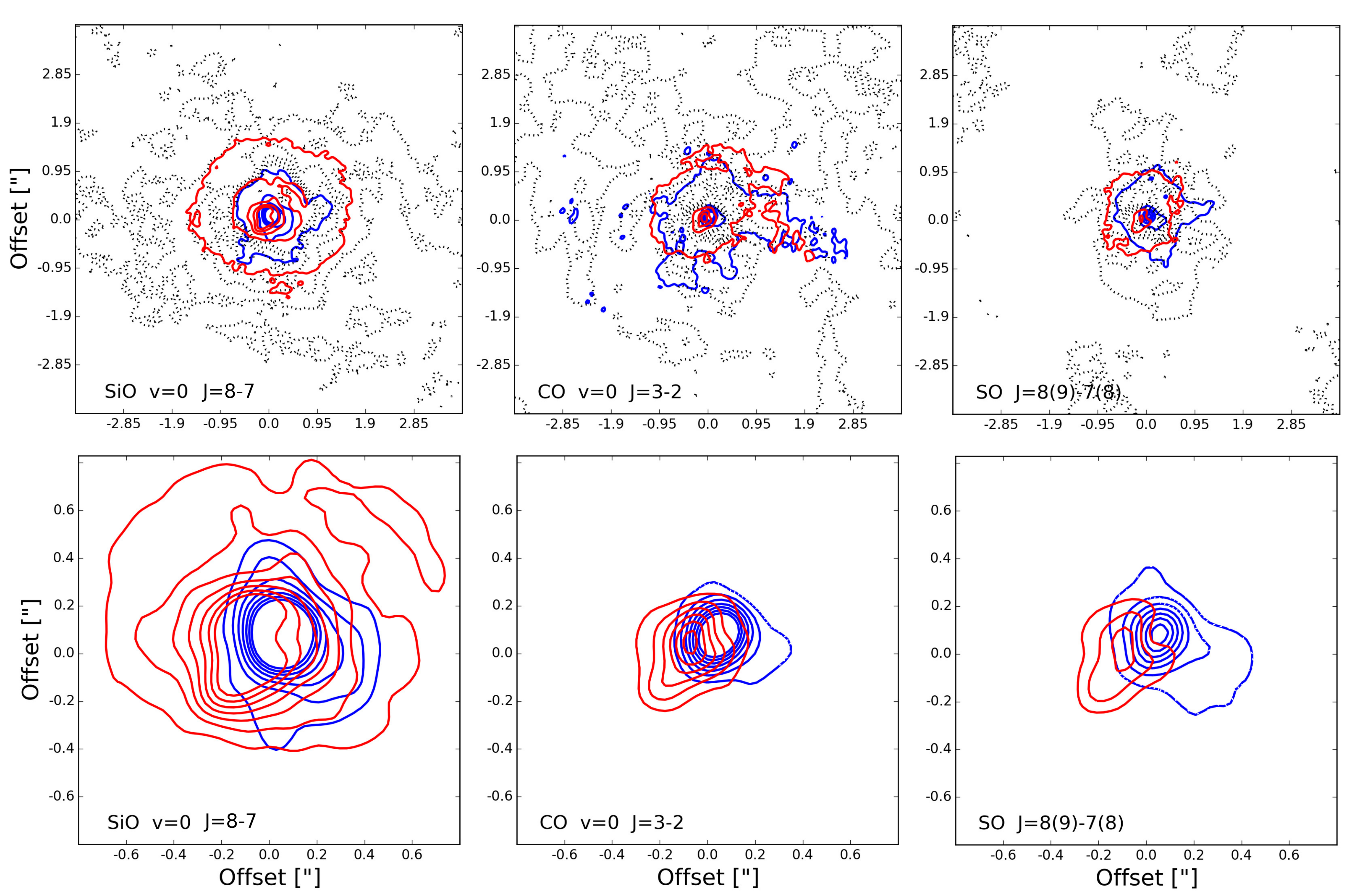}
\caption{Stereogram plots showing the offset between blue and red shifted emission of $^{\rm 30}$SiO (\textit{left}), CO (\textit{middle}), and SO (\textit{right}). North is up, west is right. The blue and red contours are constructed from 49\% of the blue and red wings of the molecular lines, respectively. The black dashed contours are constructed using the remaining 2\% of the velocity channels, centred on the $v_{\rm lsr}$. Offset is measured with respect to the peak continuum brightness position. The contours in the top (bottom) panels are evenly spaced between 3 and 51 (15 and 51) times the rms noise value outside the line ($\sim3.5\times10^{-3}$ Jy/beam). The black contours in the bottom panels have been omitted. }
\label{stereograms}
\end{figure*}

With this work we aim to extend the observational sample of AGB morphologies. Specifically, we investigate a peculiar feature seen in various molecular transitions of ALMA (cycle 2) Band 7 observations of \object{R Doradus} (hereafter R~Dor), which suggest the presence of a compact rotating disk in its inner wind. In addition, we identify a peculiar unresolved high-velocity emission feature whose chemical properties seem to differ from those of the AGB wind.

R~Doradus is a semi-regular pulsating variable \citep[$P=332-175$\,days;][]{Bedding1998} with an effective temperature of $T_\mathrm{eff} \approx 2500$\,K and a luminosity $L_* \approx 4500\,\lso$. It has a radial velocity relative to the Local Standard of Rest (lsr) of $v_{\rm lsr} \sim 7.5 \kms$. At a distance of only 59 pc \citep{Knapp2003}, R~Dor is the nearest AGB star to our Sun. This has made the star a prime target for the detailed study of the physical, chemical and dynamical properties of its oxygen-rich circumstellar environment, which results from a slow-moving wind with a terminal velocity of only $\sim 5.5\,\kms$, stripping mass from its star at a rate of $\sim 1 \times 10^{-7}\,\msoy$ \citep{VanDeSande2017}. Recent ALMA observations of R~Dor, however, show that its dynamics may be somewhat more complex than previously thought, with spectral lines showing widths up to 23$\kms$ \citep{Decin2018}. Based on its $^{\rm 17}$O/$^{\rm 18}$O abundance ratio the initial mass of R~Dor can be estimted to be $\sim$1 $\mso$\citep{DeNutte2017}.

This paper is organised as follows. In Sect. \ref{obs} we present the observations, focusing mainly on the spatial offset between the blue and red shifted emission in the inner regions of the wind. Subsequently, in Sect. \ref{model}, we reproduce the observed emission patterns through three-dimensional radiative transfer assuming a disk model. And finally, in Sect. \ref{discus}, we compare the object to the available literature on the AGB star L$_{\rm 2}$ Puppis, which features a compact rotating disk along its equator, potentially harbouring a planetary companion \citep{Kervella2016,Homan2017}. For R~Dor we too identify a peculiar feature that may point to the presence of a companion.

%__________________________________________________________________

\section{ALMA data} \label{obs}

\subsection{Observations}

R~Doradus was observed with ALMA in Band 7 in August and September 2015 (proposal 2013.1.00166.S, PI L. Decin). A full spectral scan between 335–362 GHz was made using four separate observations. An average of 39 antennas were used. The observations took place over 5 days using standard ALMA procedures. The range of baseline lengths was 0.04–1.6 km, allowing imaging of structures on angular scales up to 200 mas at angular resolution $\sim$150 mas. Each observation used four 1.875 GHz wide spectral windows.

All data reduction was done using {\tt CASA} \citep{McMullin2007}, following standard scripts for the application of calibration from instrumental measurements such as system temperature and water vapour radiometry, and using standard sources for bandpass and flux scale calibration. A detailed overview of the observation and reduction specificities are provided in \citet{Decin2018} where the emission of a wide range of molecular line transitions and the continuum emission are presented (in terms of spectral lines, channel maps, moment-zero maps and azimuthally averaged profiles). Here we present a small portion of the complete dataset that specifically relates to a particular morphological feature: a spatial offset, relative to the continuum peak, between the red and blue shifted molecular emission. 

\subsection{Offset between red and blue shifted emisson}

Upon close examination of the ALMA data, we found an offset between the red and blue shifted molecular emission across several molecular species and transitions, relative to the continuum peak, and confined to the centermost 1'' $\times$ 1'' portion of the field of view. This is seen most clearly if we separate the channels on the blue shifted side of the lsr velocity from the red shifted channels for a given transition, sum the respective emission separately and overplot the results. We henceforth refer to this type of data-visualisation as stereograms. Fig. \ref{stereograms} shows the stereograms for $^{\rm 30}$SiO ($v=0,~J=8\to7$), CO ($J=3\to2$), and SO ($N_J = 8_9\to7_8$). We show both a view of the full extent of the emission and a zoomed in view of the innermost regions where the offset is clearest. The spectral line shapes can be found in Fig. \ref{lines}.

% \begin{figure}[htp]
% \centering
% \includegraphics[width=0.4\textwidth]{SiO_pv_1pix_wl-eps-converted-to.pdf}
% \caption{}
% \label{pvwl}
% \end{figure}

\begin{figure*}[htp]
\centering
\includegraphics[width=0.4\textwidth]{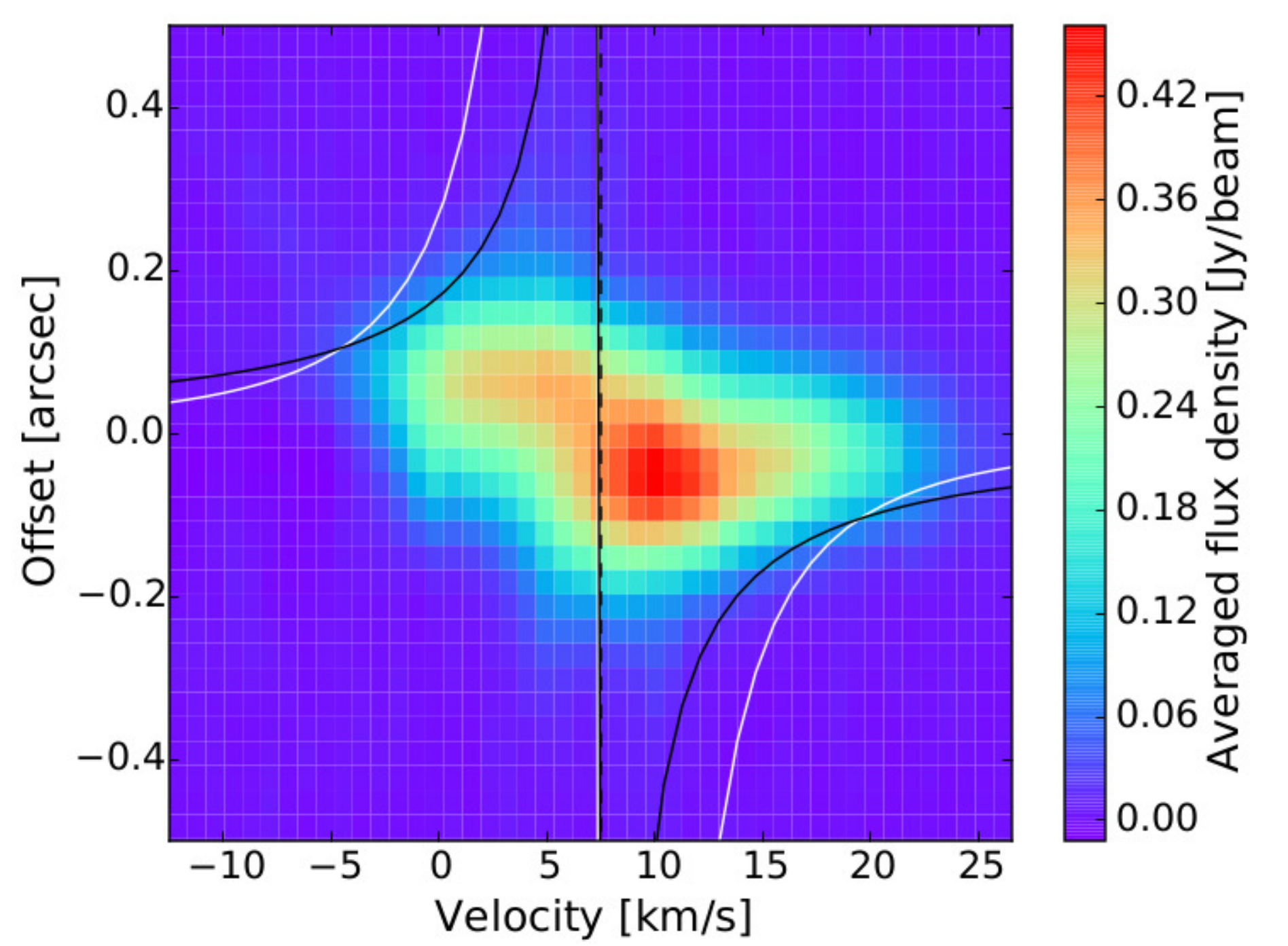}
\includegraphics[width=0.42\textwidth]{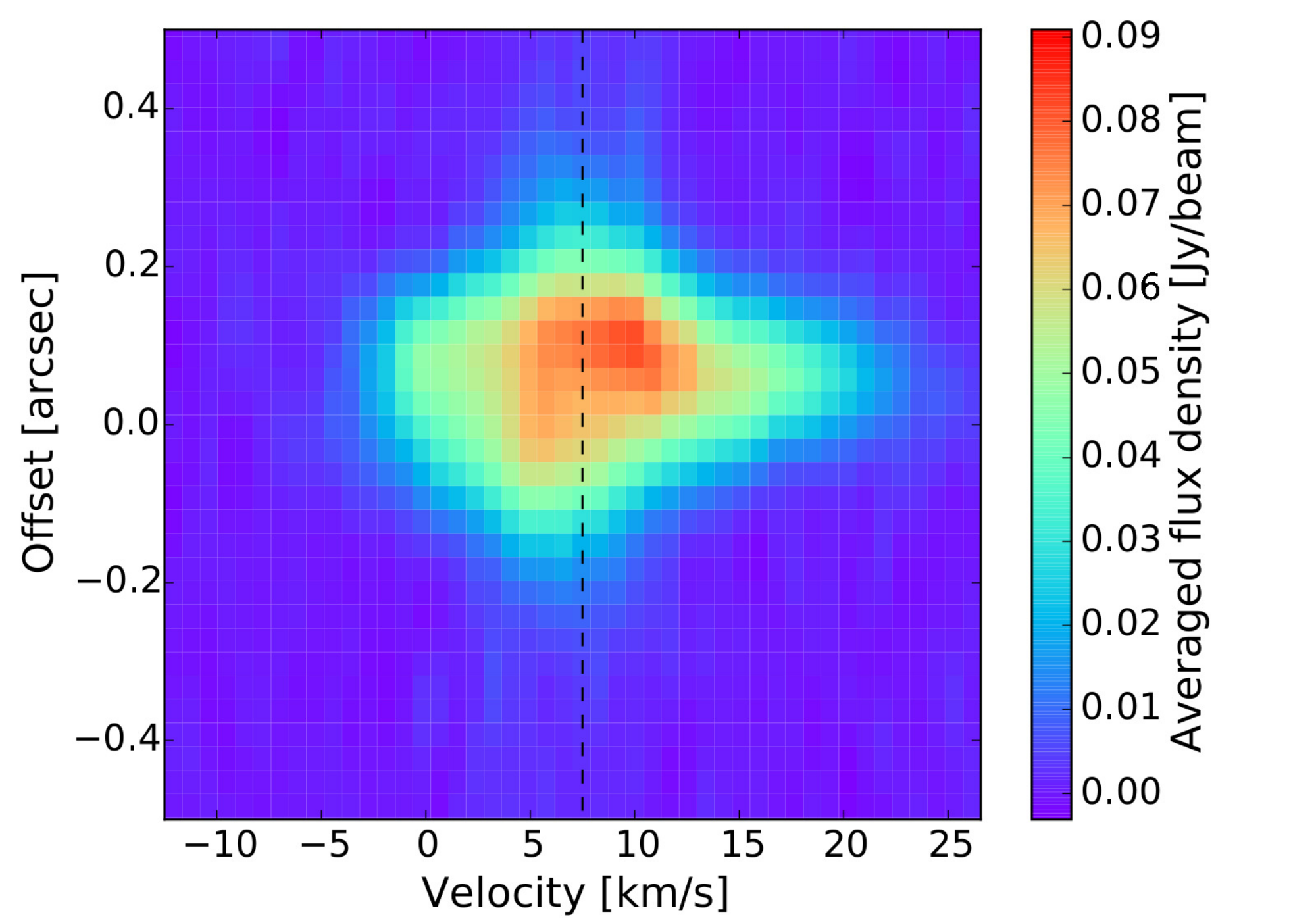}
\caption{\emph{Left panel}: Observed PV diagram of the $^{\rm 28}$SiO ($v=1,~J=8\to7$) transition, made at a position angle of $115^\circ$, with a slit width equal to 1 beam width. The curves represent different rotational regimes: the white curve illustrates a Keplerian trend, the black curve illustrates conservation of angular momentum. \emph{Right panel}: Observed PV diagrams of the $^{\rm 28}$SiO ($v=1,~J=8\to7$) transition, made at a position angle of $25^\circ$, with a slit width of 0.9''.}
\label{pvd}
\end{figure*}

\begin{figure*}[htp]
\centering
\includegraphics[width=0.4\textwidth]{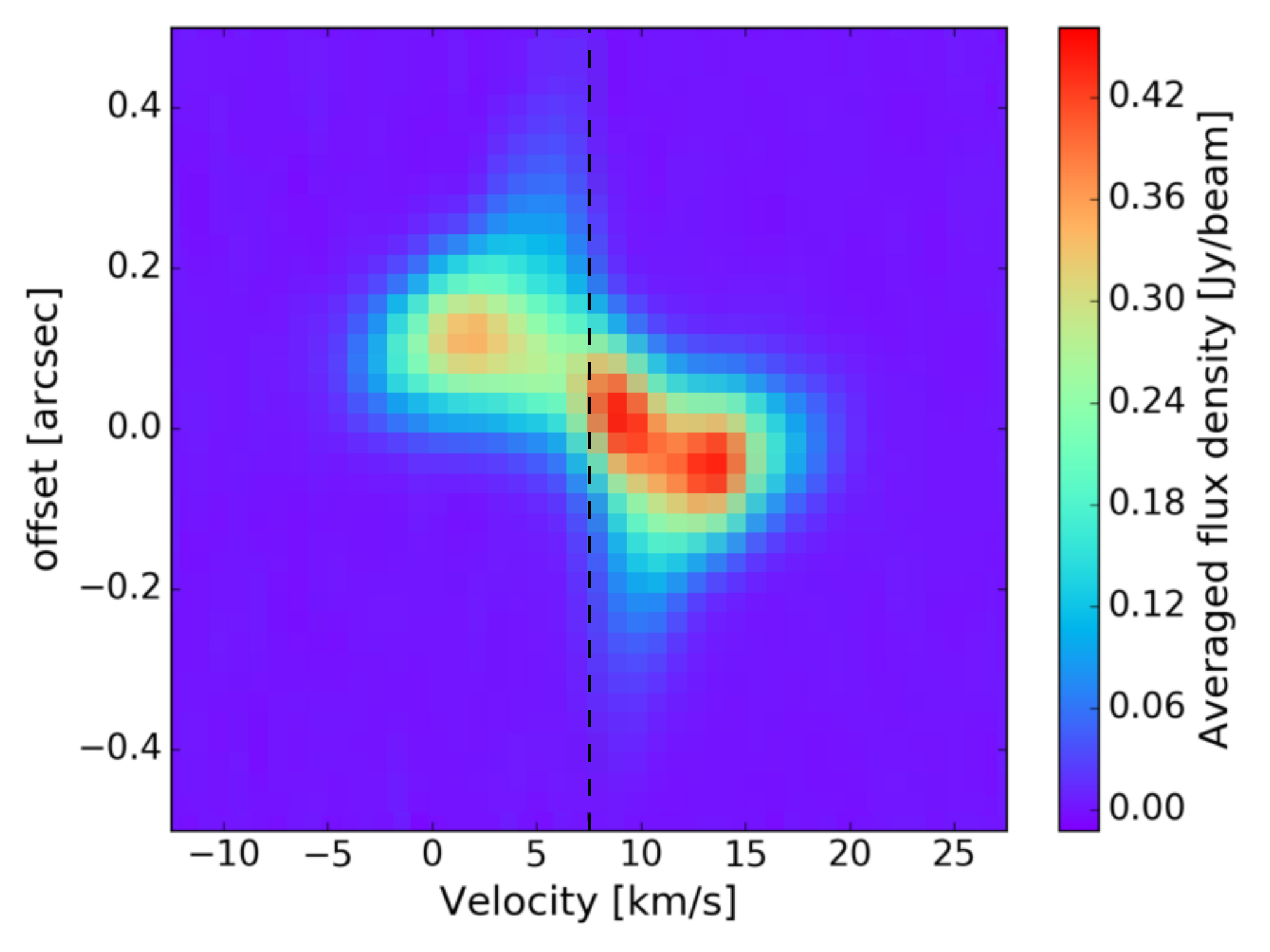}
\includegraphics[width=0.4\textwidth]{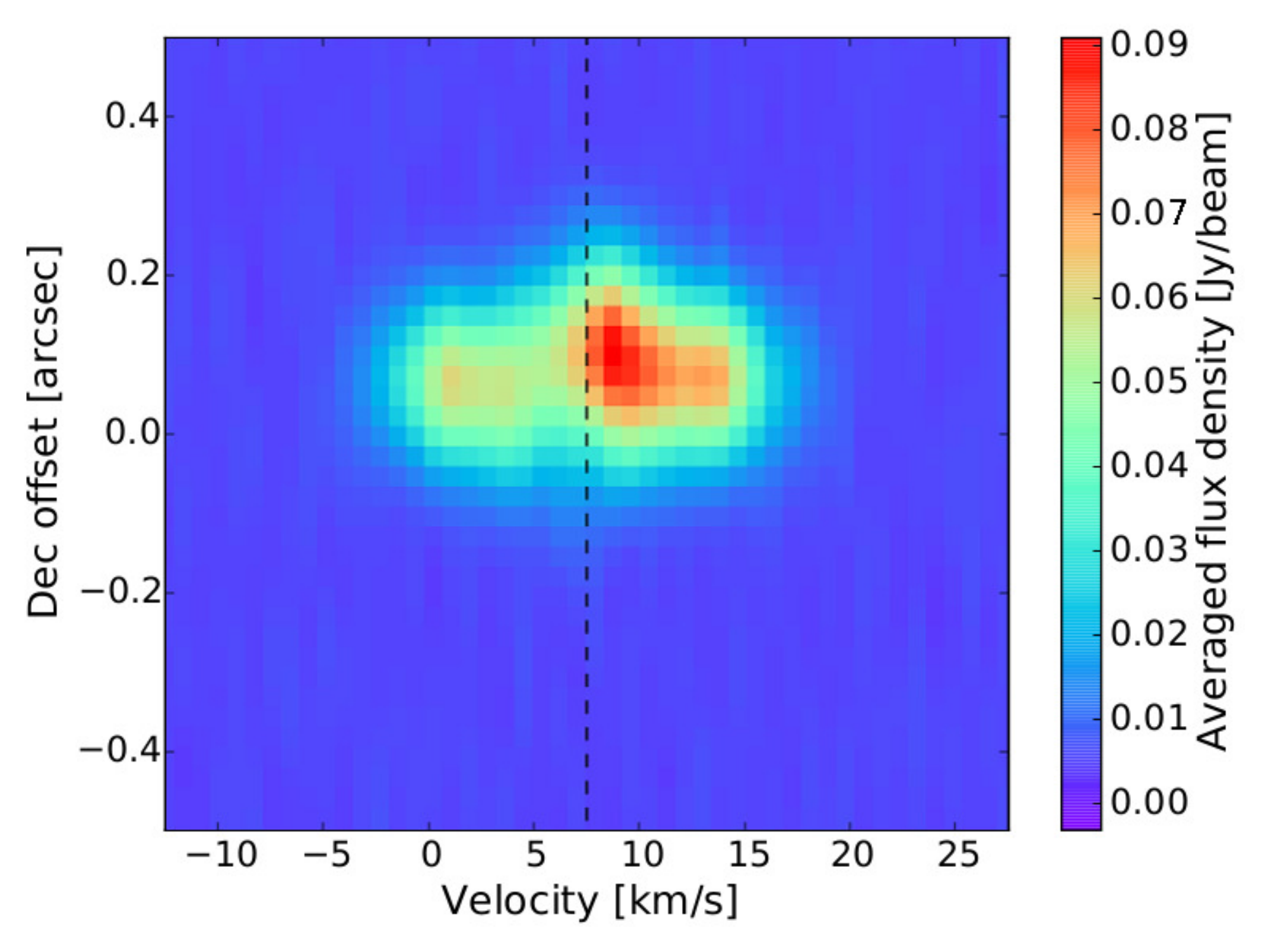}
\caption{$^{\rm 28}$SiO ($v=1,~J=8\to7$) synthetic PV diagrams of the compact disk model. \emph{Left}: PV diagram produced by placing the slit parallell to the disk equator (slit width = beam size). \emph{Right}: PV diagram obtained by placing the slit along the polar axis (slit width = 0.9'').}
\label{pvm}
\end{figure*}

From the plots in Fig. \ref{stereograms}, the axis of the offset appears to be close to the east-west axis. To more accurately determine the position angle of the system we made orthogonal pairs of position-velocity (PV) diagrams with axes at $5^\circ$ intervals through the centre of the image. We chose to focus the analysis on the $^{\rm 28}$SiO ($v=1,~J=8\to7$) transition which is mainly excited in the centermost regions and does not exhibit extended emission that would otherwise severely contaminate the signal. The channel maps of this transition are shown in Fig. \ref{chan}. We determine that the geometrical symmetry axes of the signal lie around a position angle (PA) of $25^\circ$ (as measured counterclockwise from north in Fig. \ref{stereograms}) and around $115^\circ$. The uncertainty on these estimates is $\sim5^\circ$.

Shown in Fig. \ref{pvd} (left panel) is the PV diagram along the $115^\circ$ axis, with a slit width equalling one ALMA resolution element. This shows the spectral evolution of the disk emission along the axis perpendicular to the projected polar axis of the tentative disk. Since the angular size of the $^{\rm 28}$SiO signal is only approximately twice the size of one ALMA resolution element (see Fig. \ref{chan}), no fine detail is visible in the diagram. However, a global overall shape of the signal in the PV-plane is clearly visible. The PV diagram shows a point-reflection symmetry around the (0,0) point (after correcting the velocty for the $v_{\rm lsr}$ offset), with a clear symmetrical offset inversion between the blue and red shifted emission. Such signals have been recovered in the PV diagrams of evolved stars with confirmed semi-Keplerian circumstellar dynamics \citep[e.g.][]{Bujarrabal2015,Bujarrabal2016}, tracing a disk along the equatorial axis. A brightness asymmetry is seen between the west and east wing that could be explained by the presence of a small radial velocity component in the disk. We also observe a so far unexplained increased red-shifted velocity of the PV signal.

We also constructed an orthogonal wide-slit (0.9'') PV diagram, shown in the right panel of Fig. \ref{pvd}. It shows a central, almost rhomboidal appearance. Combined with its orthogonal partner, these shapes are suggestive of the presence of an inclined differentially rotating disk-like structure, as shown by \citet{Homan2016}.

A disk hypothesis seems to be in agreement with the recent polarimetric continuum observations of the inner wind of R~Dor with SPHERE \citep{Khouri2016}. The infrared images presenting the degree of polarisation in different bands in Fig. 5 in Khouri et al. exhibit a pattern that strongly resembles the `butterfly' shape seen in L$_{\rm 2}$ Puppis by \citet[][ Fig. 4]{Kervella2015}. This pattern has been interpreted as the scattering of the incident stellar light onto the inner rim of a dust disk, which, for a nearly edge-on orientation, lies almost perpendicular to the line of sight. For R~Dor, the pattern is located at a distance of 50 - 100 mas from the star, placing the dust disk's inner rim at an absolute distance of 3 - 6 AU from the central star.

\begin{table*}[htp]
\centering
\caption{R~Dor disk model density and geometrical parameters, versus L$_{\rm 2}$ Puppis. \label{dens}}
\begin{tabular}{ l r r }

\hline
\hline
Disk Density Parameters & & \\
\hline
 & R~Doradus & L$_{\rm 2}$ Puppis \\
\hline
Inner rim $r_{\rm c}$ & $6.0 {\rm\ AU}$ & $2.0 {\rm\ AU}$ \\
Outer radius & $\sim 25 {\rm\ AU}$ & $\sim 30 {\rm\ AU}$ \\
Scale height at inner rim $H_c$ & $0.9 {\rm\ AU}$ & $1.5 {\rm\ AU}$ \\
Flaring factor $h$ & $0.20$ & $0.20$ \\
Density at inner rim $\rho_0$ & $3.5 \times 10^{-12} {\rm\ kg/m}^3$ & $9.3 \times 10^{-10} {\rm\ kg/m}^3$ \\
Radial density power index $p$ & $-2.8$ & $-3$ \\ 
Inclination angle & $110^\circ$ & $80^\circ$ \\
\hline

\end{tabular}
\end{table*}

\section{Radiative transfer model and model assumptions} \label{model}

Using the 3D non-local-thermodynamical-equilibrium (NLTE) radiative transfer code {\tt LIME} \citep{Brinch2010}, in combination with the observation simulation algorithms of {\tt CASA} \citep{McMullin2007}, we aim to reproduce the observed three-dimensional emission distribution by assuming a differentially rotating disk. Following \citet{Homan2017}, its density is described by the analytical solution for a thin, non-self-gravitating, vertically isothermal disk in hydrostatic equilibrium:

\begin{equation}
 \rho(r_{xy},\phi,z) = \rho_0\left(\frac{r_{xy}}{r_{\rm c}}\right)^{p}\exp\left[\frac{-z^2}{2H(r_{xy})^2}\right], 
\end{equation}
with $(r_{xy},\phi,z)$ the standard cylindrical coordinates, $r_{\rm c}$ the inner rim of the disk (in the equatorial plane), $\rho_0$ the H$_{\rm 2}$ gas density at $r_{\rm c}$, $p$ a measure for the rate with which the density subsides radially, and $H(r_{xy})$ the vertical Gaussian scale height of the disk, expressed as
\begin{equation}
 H(r_{xy}) = H_c\left(\frac{r_{xy}}{r_{\rm c}}\right)^h,
\end{equation}
where $H_c$ is the initial scale height at the location of the inner rim in the midplane of the disk, and $h$ a measure for the rate at which the scale height increases radially. The derived density parameter values are listed in Table \ref{dens}.

The disk temperature is assumed to be a simple power-law with an index of $-$0.5. In order to reproduce the kinematical behaviour displayed in Fig. \ref{pvd} the velocity field of the disk needs to be sub-Keplerian, and approximately follows conservation of angular momentum ($v \sim 1/r$), starting with Keplerian motion at the inner rim location. We measure the maximal velocity of the PV signal of the $^{\rm 28}$SiO emission as $\sim$12 $\kms$, which, for a 1 $\mso$ central star, corresponds to an inner rim location of $\sim$6 AU. The sub-Keplerian nature of the rotation profile is illustrated by the solid black and white curves in Fig. \ref{pvd}. To model the asymmetry in the emission we add a radial velocity component to the rotation profile, with an outward pointing magnitude of 1$\,\kms$. This type of decretion is to be expected in sub-Keplerian regimes due to the radial force imbalance. The turbulent velocity component is 1 $\kms$. The molecular abundance of [SiO/H$_{\rm 2}$] = 6$\times$10$^{\rm -5}$ has been taken from \citet{VanDeSande2017}. Because we cannot get accurate estimates of the height-to-width ratio, we cannot constrain the inclination of the system. However, by comparing the shapes of the orthogonal PV diagrams with the theoretical predictions in \citet{Homan2016}, we estimate the disk to be inclined at an angle of $\sim$110$^\circ$ away from the face-on position, with the side closest to the observer positioned above the line of sight through the disk center. The height of the disk is the result of a combined effect of the $H_c$ and $h$ parameters, which cannot be properly determined individually because of the relatively large beam size of the observation. The assumed values are rougly based on the results of L$_{\rm 2}$ Puppis \citep{Homan2017}. The disk is assumed to surround a central star emitting as a black body with a temperature of 2500 K and with a mass of 1 $\mso$ \citep{Danilovich2017}. For the SiO atomic model we adopt the spectroscopic data in the LAMDA database \citep{Schoier2005}, with 61 rotational levels in each of the three lowest vibrational levels; the collisional rates are taken from \citet{Yang2010}. The contribution of the dust emission to the mean intensity of radiation was modelled using the results of \citet{VanDeSande2017}, with an assumed gas-to-dust mass ratio of 100. The external observational conditions, as well as the specific instrumental set-up at the time of observation have been used as input for the {\tt CASA} observation simulation algorithms. 

The higher excitation vibrational lines of SiO often show master emission. The moderate brightness of the considered line suggests that SiO is not masering in the considered transition. However, to be certain we checked the level populations calculated by {\tt LIME} to identify whether population inversion has not occurred. We found no indication that the optical depth $\tau$ has become negative at any position or iteration during the radiative transfer calculations. 

\subsection{Model results}

The model parameters can be found in Table \ref{dens}. The PV diagrams of the resulting emission distribution are shown in Fig. \ref{pvm}. Both the global shape as well as the absolute extent and emission of the features are reasonably well reproduced by our disk model. Some important differences remain, though. The central emission in the synthetic wide-slit PV diagram is a little brighter and has a more defined shape compared to the data. In addition, the data seems somewhat more fuzzy and exhibits a slightly increased red wing that can not be explained by the current symmetrical model. Finally, the width of the synthetic signal along the diagonal axis seems to underestimate the data. These differences probably reflect that the geometry, density and temperature distribution is more complex than captured by our simplifying assumptions, in addition to the complexity of the medium in which the tentative disk is embedded. Nevertheless, the overall behaviour is reproduced by the model.

We note that the results obtained here are the outcome of a carefully chosen set of parameters whose values have been varied until a reasonable visual resemblance between the model and the data was achieved. Hence, we can make no statements about the quality of the `fit'. Nevertheless, we attribute the next section to a qualitative description of the final model sensitivity to the parameter values.

Having estimated the disk density, we can calculate its mass (to first order) by integrating over the volume. We recover a disk mass amounting to approximately $6 \times 10^{24} \rm kg$ or $3 \times 10^{-6} \mso$. Assuming the dynamics of the disk conserve angular momentum, we can also estimate the total angular momentum contained within the disk. This amounts to approximately $5 \times 10^{40}\ {\rm m^2 kg/s}$. 

\subsection{Parameter sensitivity}

The kinematics in the disk can in principle be determined by measuring the slope of the edge of the PV signal along the disk equator. This was done very precisely by \citet{Kervella2016}, and was attempted in a more rudimentary form in Fig. \ref{pvd}. However, the insufficient spatial resolution of the data imposes some important limitations on this analysis. We can only roughly trace the edges of the PV signal with analyical trends of differential rotation, and through visual inspection roughly esimate the velocity trend needed to reproduce this. We conclude that the edge of the PV signal can be explained by assuming sub-Keplerian rotation. The assumed velocity field is thus probably a decent first order approximation, but may require significant correction once higher spatial resolution images of the object are made.

The spectral resolution of the data is high enough to pinpoint the location of the disk's inner rim relatively well. The highest measured projected velocities for an almost edge-on disk orientation give an inner rim location of $\sim$6 AU. However, this is based on the emission of a highly radiatively excited molecule prone to masering. We should therefore be careful when interpreting the results. For example the line may only be excited some distance from the actual inner rim of the disk. The SPHERE data of R~Doradus \citep{Khouri2016} indicate that the disk's inner rim is probably located between 3 and 6 AU, in agreement with with the data presented here.

The height of the disk can be estimated from the height of the PV signal along the disk meridian. We manage to reproduce it relatively well assuming an inclination angle of $110^\circ$. We can not properly determine the inclination of the system, so we assume an uncertainty of 20$^\circ$ on the inclination angle. This results in an uncertainty estimate on the disk height of $\sim$60\%.

The most important assumption which permits us to deduce the disk density properties is its temperature profile. There is currently no way to gauge whether the adopted temperature profile is valid, and hence any uncertainties here translate directly to uncertainties on all derived density parameters. The density at the disk's inner rim was estimated by attempting to reproduce the highest absolute emission values inside the PV diagrams. Assuming no uncertainty on the temperature, we estimate an uncertainty of $\sim$30\%. Finally, the radial density power slope was estimated based on the maximum offset of the PV signal (along the disk equator). The extent of this signal is rather sensitive to the slope. Assuming again no uncertainty on the temperature, we estimate an uncertainty on the slope of $\sim$15\% or $p=-3\pm0.5$.

\section{Discussion} \label{discus}

\subsection{Comparison to L$_{\rm 2}$ Puppis}

Recent models of the circumstellar environment of the AGB star L$_{\rm 2}$ Puppis report the first detection of a differentially rotating disk around an AGB star \citep{Kervella2016}, and provide an in-depth view of its geometrical and (thermo)dynamical structure \citep{Homan2017} by modelling its CO emission. We find here that the main properties of the tentative disk around R~Dor are rather similar to those of the disk around L$_{\rm 2}$ Pup, as shown in Table \ref{dens}. The L$_{\rm 2}$ Pup disk is an extremely compact equatorial density enhancement with a scale height of approximately 5 AU and a radius of approximately 25 AU. We could not precisely map the dynamics in the R~Dor disk. Additionally, we could not identify any radial changes in dynamical regime (like a sharp kinematical transition between Keplerian and sub-Keplerian rotation at $\sim$5-6 AU detected by Kervella et al. 2016). Hence, we have assumed the disk to be fully sub-Keplerian. Interestingly, the steep radial power index $p$ of $\sim-$3 for the density of the disk found in L$_{\rm 2}$ Pup is also recovered for the R~Dor system to explain the equatorial extent of the signal.

Both the mass and angular momentum of the disk are much lower than in the L$_{\rm 2}$ Pup case. This is directly linked to the hence most important morphological difference: the location of the inner rim. For R~Dor we place the inner rim at $\sim$6 AU, while for L$_{\rm 2}$ Pup it is at $\sim$2 AU. The R~Dor inner rim radius is consistent with SPHERE polarimetric imaging \citet{Khouri2016}, which indicates an inner rim location between 3 and 6 AU. The determination of the inner rim location may be hampered by the resolution and sensitivity of the observation, which may cause discontinuities in the weak $^{\rm 28}$SiO ($v=1,~J=8\to7$) signal of the inner disk at high velocities, making the PV feature appear more confined in velocity-space than it actually is. In addition, L$_{\rm 2}$ Pup shows an important stratification in the dusty component of the disk. This cannot be verified properly with the current data for R~Dor, though the sub-Keplerian nature of the disk may point to radiation pressure on its dusty contents as a factor determining the disk velocity structure, as is the case for L$_{\rm 2}$ Pup \citep{Haworth2017}.

Following \citet{Homan2017} we estimated the total angular momentum within the central star to be $\sim10^{41}\ {\rm m^2 kg/s}$. This amounts to about twice the total angular momentum in the disk. However, the mass of the disk is only a negligible fraction of the total mass lost by the star. It is thus highly unlikely for this amount of angular momentum to have been transferred from the star into the disk via simple mass transfer. This classical angular momentum problem can be solved, in analogy with the analysis of the L$_{\rm 2}$ Pup ALMA data \citep{Homan2017}, by assuming the presence of a (planetary) companion of significant mass at the location of the disk's inner rim. We note that in our own solar system Jupiter contains $\sim$99\% of the system's angular momentum. This planetary companion serves as an angular momentum injector into the local CSE. Assuming an angular momentum transfer efficiecy of $\sim$30\% \citep{Nordhaus2006,Nordhaus2007}, the angular momentum in the disk could be explained by a companion with a mass of at least $1.5 \times 10^{25} \rm kg$ or only about 2.5 earth masses at a distance of 6 AU. We note that this is a preliminary lower limit estimate, and that follow-up observations with ultra-high spatial resolution are needed to refine both the radiative transfer model and the estimates of the system properties.

\subsection{Tentative detection of gas associated with a planetary companion}

Several molecules exhibit a peculiar high-velocity feature in the extreme blue wing of their spectral lines, broadening them significantly beyond the expected width, as discussed in \citet{Decin2018}. The broadening is caused by a faint patch of unresolved emission (with an area comparable to the beam size), located in the south-east, at a distance of $\sim$0.2'' from the central star. We refer to this feature henceforth as the blue blob. It is present in the channels between 8 and 18 $\kms$, giving it an average speed of $\sim$ 13 $\kms$ relative to the $v_{\rm lsr}$ (see Fig. \ref{SO2chan}). 

Its extreme speed with respect to previously measured outflow velocity of only 5.5 $\kms$ at the observed (projected) distance renders it unlikely that it is associated with the AGB wind. The feature has no counterpart in thermal continuum emission, which too may be an indication that it is not part of the stellar wind as dust grains would likely have formed in such a dense environment if the structure was indeed a clumpy part of the AGB-outflow. The non-detection in continuum light is likely not caused by a lack of sensitivity, as the current noise level in the continuum (40 $\mu$Jy) is comparable to the L$_{\rm 2}$ Pup observation (sensitivity 34 $\mu$Jy), where a prominent continuum source (located at the disk's inner rim) is also found to exhibit molecular emission. Intriguingly, the blue blob is seen in different degrees of brightness in all observed molecules, except HCN (see Fig. \ref{HCNchan}). Though HCN emission is strong throughout in the CSE of R~Dor, it is, surprisingly, altogether lacking in the blob. From thermodynamic equilibrium arguments HCN is not expected to form in O-rich environments. Its presence in the inner wind of R~Dor may be explained by pulsation-induced non-equilibrium shock-chemistry \citep{Gobrecht2016} or by effects of photochemistry in a porous medium (Van de Sande et al. \emph{subm.}). 

Given the peculiar properties of the blue blob we speculate that it may be associated with a physical object embedded in the outflowing wind that created a local environment where neither shocks nor photons dominate the local chemistry. The mean velocity of the blob coincides rather well with the deduced maximal velocity of the disk, i.e. the tangential velocity at the 6 AU inner rim (12 $\kms$). This is similar to the recent detection of the tentative planetary companion at the inner rim of the disk surrounding the AGB star L$_{\rm 2}$ Pup. However, the scenarios are not identical. In the L$_{\rm 2}$ Pup case, the tentative companion manifested itself strongly in continuum light, and faintly in molecular emission \citep{Kervella2016,Homan2017}. Here, we do not detect the blue blob in the continuum, but it is present in the emission of most observed molecules. Given the indications discussed, we hypothesise that the blue blob may be a feature linked to a companion planet to R~Dor. In this case the velocity at which the material is seen under the inclination angle of $110^\circ$ would correspond to a Keplerian orbit at a projected distance of $\sim$ 0.05'' from the central star. This is  smaller than the current measured peak distance of 0.2''. However, note that the emission is fully unresolved. Possibly, the proximity of the potential planet to the central star has caused it to puff up, causing its loosely bound outer layers to be stripped by the wind of R~Dor. This scenario could explain the difference in the inferred position of the planetary companion and the location of the peak emission, the latter originating in a \emph{tail} of shed material.

A multitude of different assumptions need validation to support this hypothesis. Firstly, we assume the mean speed of the accelerated wind to be $\sim$5.5$\kms$. But, wind speed measurement by \citet{Decin2018} indicate that a mechanism may accelerate AGB winds far beyond previously measured outflow velocities. There is currently no explanation for this phenomenon. An increased terminal wind speed would imply that the blob may indeed be part of the wind, in which case it would have to be a chemically dissimilar clump of gas. Furthermore, it would have to be very small and directionally ejected by the AGB star. The current quality of the data does not permit us to confirm or exclude any hypothesis regarding the nature and origin of the blue blob. Secondly, we assumed the disk to exhibit Keplerian speeds at the location of the tentatively determined inner rim. As the analysis of the L$_{\rm 2}$ Puppis data has shown, this need not be the case. The presence of dust in the disk can exert an outward pressure force on the gas that could reduce the effective gravity on the disk material, effectively reducing its rotation speed \citep{Haworth2017}. And seeing as the SPHERE data indicates that the inner rim of the dust disk can range between 3 and 6 AU, reduced tangential speeds at 3 AU could resemble Keplerian speeds at 6 AU, and thus also be able to explain the data. Finally, we compare the mean velocity of the blue blob with the tangential speeds at the inner rim, which does not contradict the plausibility of the hypothesis. However, the margins of uncertainty are substantial. The velocity width of the blue blob signal, and the uncertainty associated with its real position relative to its host star and the earth (due to projection effects) can easily conspire to break down the hypothesis. We thus conclude that follow-up high-resolution observations are needed to test our claims, and to deeper investigate the true nature of both the disk-like signal and the blue blob.

\section{Conclusions}

A spectral scan of the circumstellar environment of the AGB star R~Dor was acquired with ALMA at a spatial resolution of $\sim$ 150 mas. Among the wealth of data, many molecular transitions show a distinct spatial offset between the blue and red shifted emission in the central 1'' $\times$ 1''. We constructed position-velocity diagrams of the $^{\rm 28}$SiO ($v=1,~J=8\to7$) transition, selected among the multiple of other transitions exhibiting this characteristic mainly because of its strong spatial confinement, mitigating any contamination by larger scale emission. Orthogonal wide-slit PV diagrams exhibit emission patterns whose dynamical signature resemble those of a compact, nearly edge-on differentially rotating disk, with a position angle of $115^\circ$. We model the emission using the radiative transfer code {\tt LIME}, and reproduce the morpho-kinematical emission features by assuming a compact disk, inclined under an angle of $110^\circ$ away from a face-on orientation. The radial density profile follows a radial power-law with an index $-$2.8, assuming a temperature drop-off proportional to $r^{-1/2}$. We calculate the mass of the disk and its angular momentum. We estimate that the angular momentum in the disk vastly exceeds any angular momentum the AGB star could have contributed, presenting a classical `angular momentum problem'. An object with a mass of at least $\sim$2.5 earth masses at a distance of 6AU, the tentative disk inner rim location, could explain the angular momentum contained within the disk, though its mass may be substantially higher.

An isolated clump of emission is detected to the south-east with a velocity that is high compared to previous terminal velocity estimates of the wind. The chemical composition of this feature seems to differ from the surrounding wind composition, as it does not show any HCN emission which is otherwise very bright in the wind. We hypothesise that its position and mean velocity may point to an (evaporating) companion planet, located at the disk's inner rim. Future ultra-high resolution observations of the system are needed to test this and reveal the true nature of both the disk and a tentative companion.

\begin{acknowledgements}

W.H. acknowledges support from the Fonds voor Wetenschappelijk Onderzoek Vlaanderen (FWO). LD acknowledge support from the ERC consolidator grant 646758 AEROSOL. TD acknowledges support from the FWO Research Project grant G024112N. This paper makes use of the following ALMA data: ADS/JAO.ALMA2013.0.00166.S. ALMA is a partner- ship of ESO (representing its member states), NSF (USA) and NINS (Japan), together with NRC (Canada) and NSC and ASIAA (Taiwan), in cooperation with the Republic of Chile. The Joint ALMA Observatory is operated by ESO, AUI/NRAO and NAOJ.

\end{acknowledgements}

\bibliographystyle{aa}
\bibliography{wardhoman_biblio}

% \clearpage
\begin{appendix}

\section{SiO ($v=1,~J=8\to7$) channel maps}

\begin{sidewaysfigure*}[]
        \centering
        \includegraphics[width=25cm]{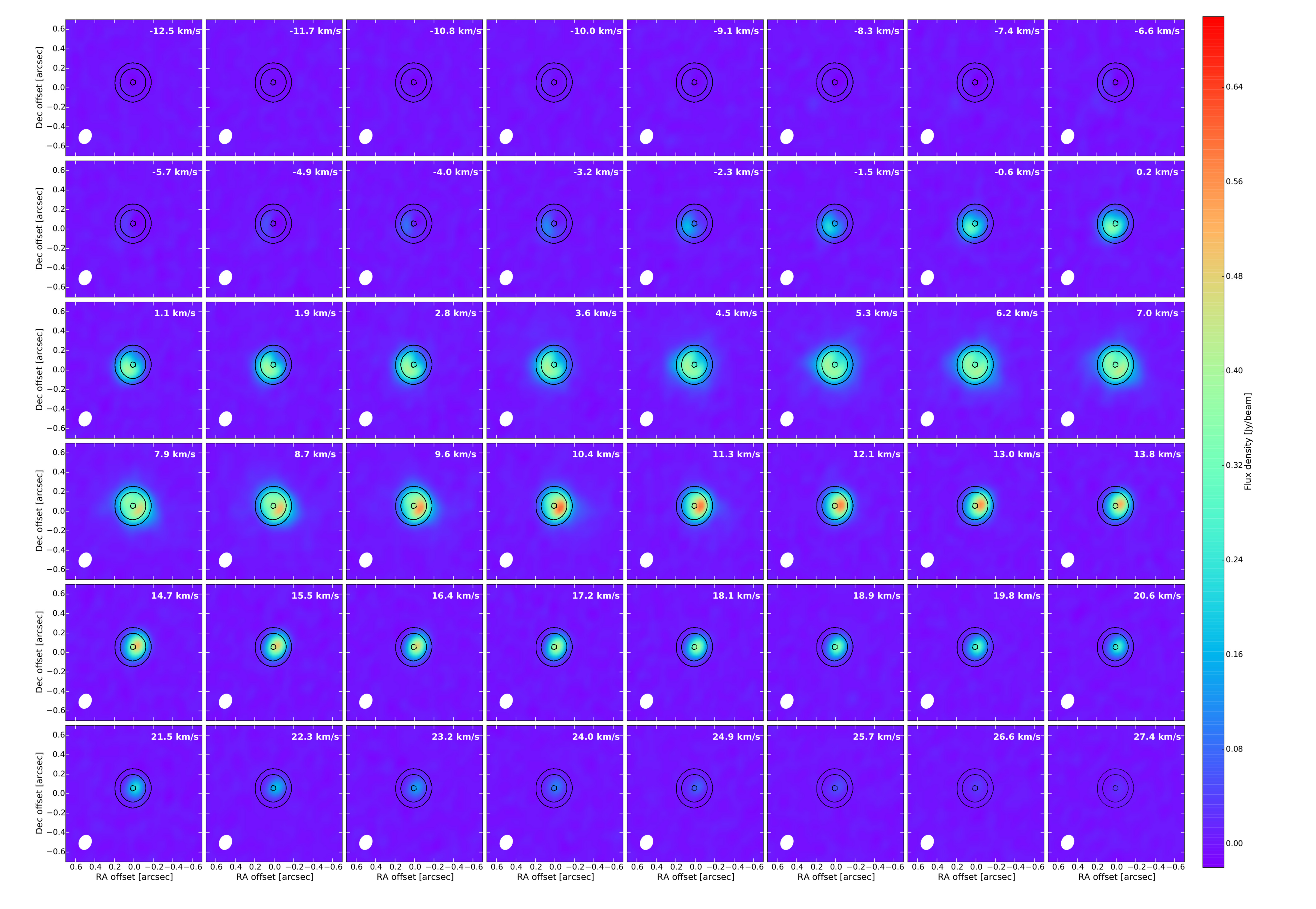}
        \caption{Channel maps of the SiO ($v=1,~J=8\to7$) emission. The black contours represent the 99\%, 10\% an 1\% contours of the peak continuum flux (0.6 Jy/beam), details on the nature of the continuum emission can be found in \citet{Decin2018}. The ALMA FWHM beam size is represented by the white ellipse in the bottom left corner.
        \label{chan}} 
\end{sidewaysfigure*}

\section{\emph{Blue blob} - SO$_{\rm 2}$ and HCN channel maps}

\begin{sidewaysfigure*}[]
        \centering
        \includegraphics[width=25cm]{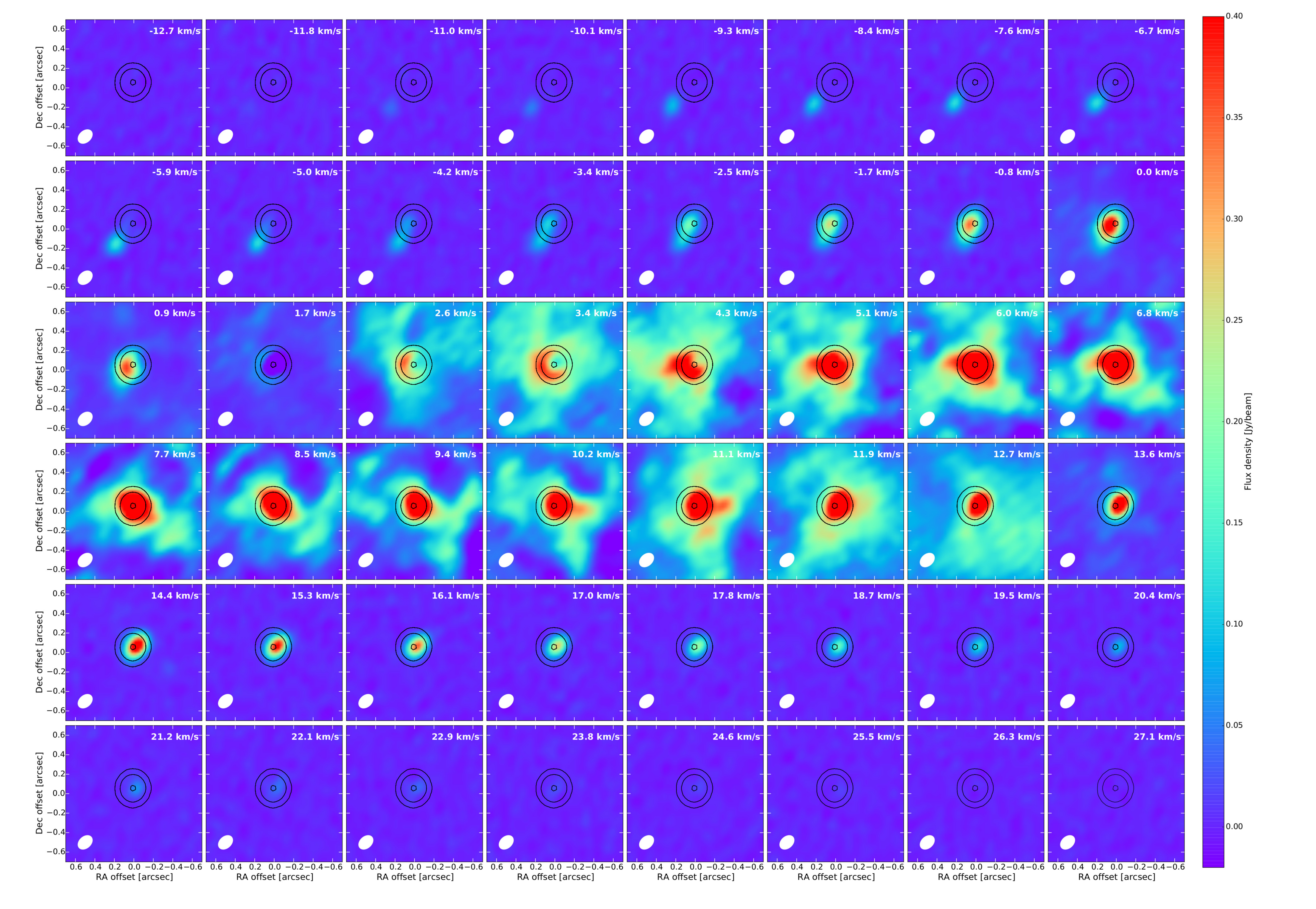}
        \caption{Channel maps of the CO ($v=0,~J=3\to2$) emission, averaged over 2 channels. The black contours represent the 99\%, 10\% an 1\% contours of the peak continuum flux (0.6 Jy/beam), details on the nature of the continuum emission can be found in \citet{Decin2018}. The ALMA FWHM beam size is represented by the white ellipse in the bottom left corner. The blue blob is visible in the channels between -10.4 $\kms$ and -1.4 $\kms$.
        \label{SO2chan}} 
\end{sidewaysfigure*}

\begin{sidewaysfigure*}[]
        \centering
        \includegraphics[width=25cm]{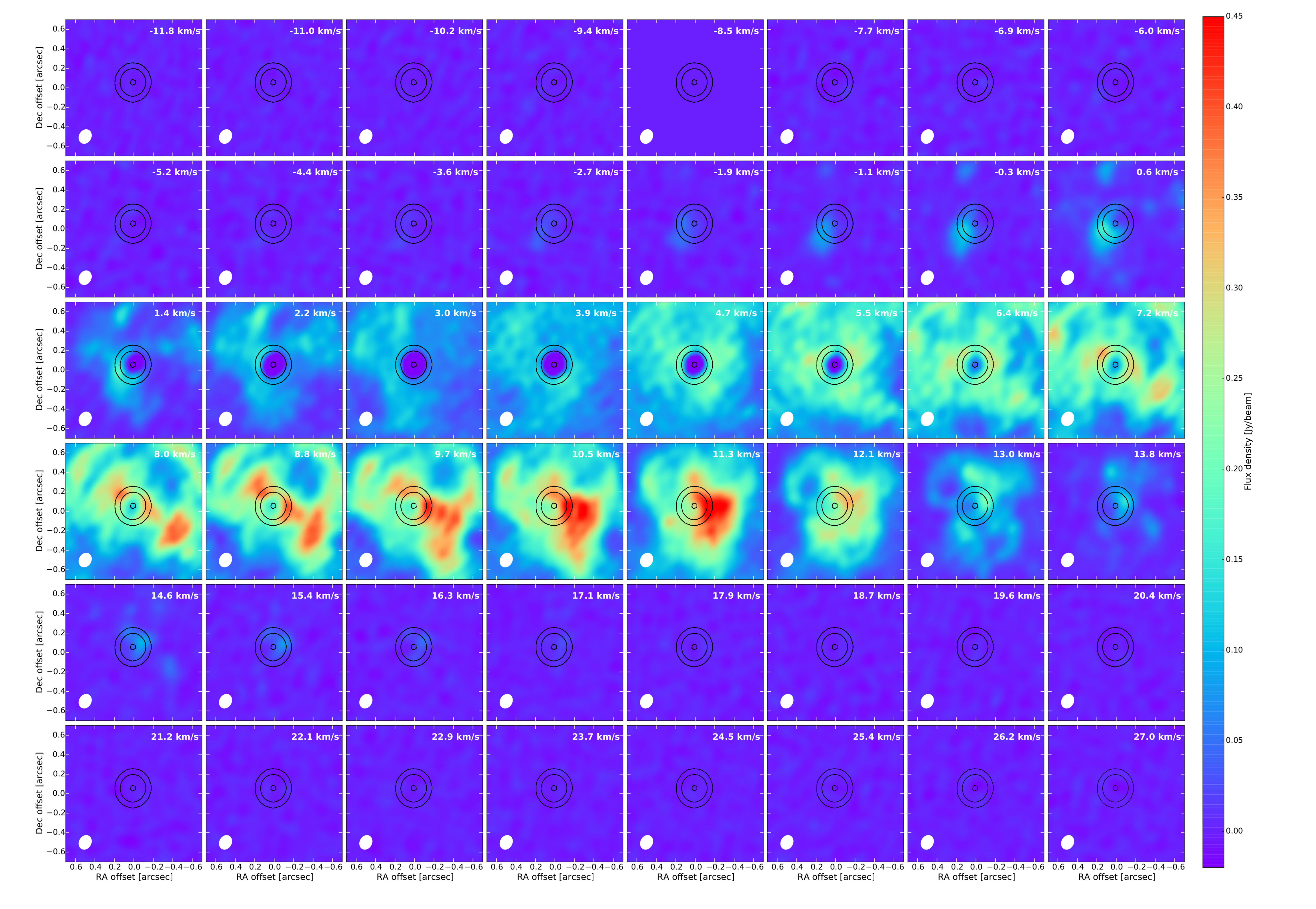}
        \caption{Channel maps of the HCN ($v=0,~J=3\to2$) emission, averaged over 2 channels. The black contours represent the 99\%, 10\% an 1\% contours of the peak continuum flux (0.6 Jy/beam), details on the nature of the continuum emission can be found in \citet{Decin2018}. The ALMA FWHM beam size is represented by the white ellipse in the bottom left corner.
        \label{HCNchan}} 
\end{sidewaysfigure*}

\section{Spectral lines of the discussed molecular emission}

\begin{figure*}[]
        \centering
        \includegraphics[width=9cm]{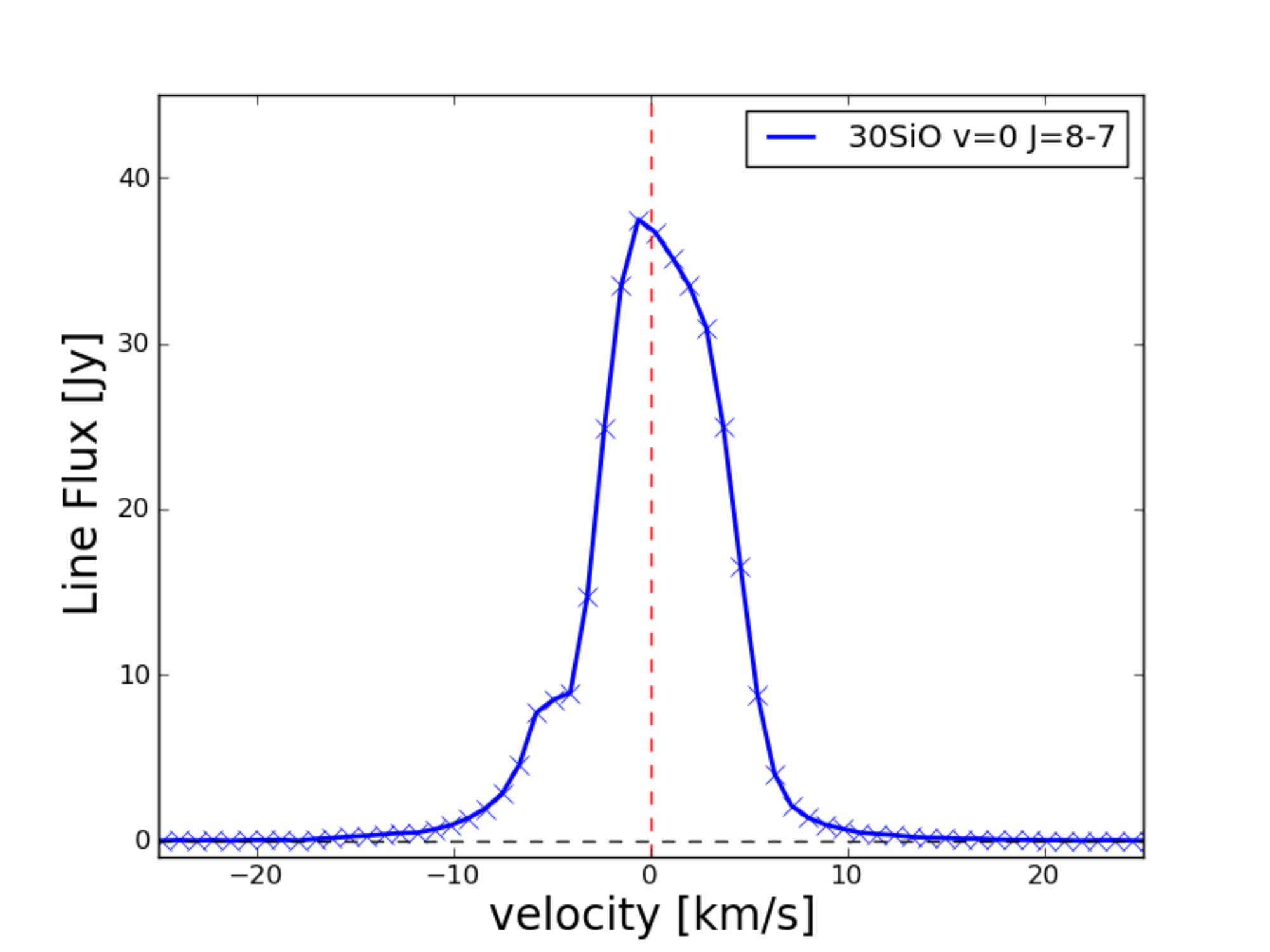}
        \includegraphics[width=9cm]{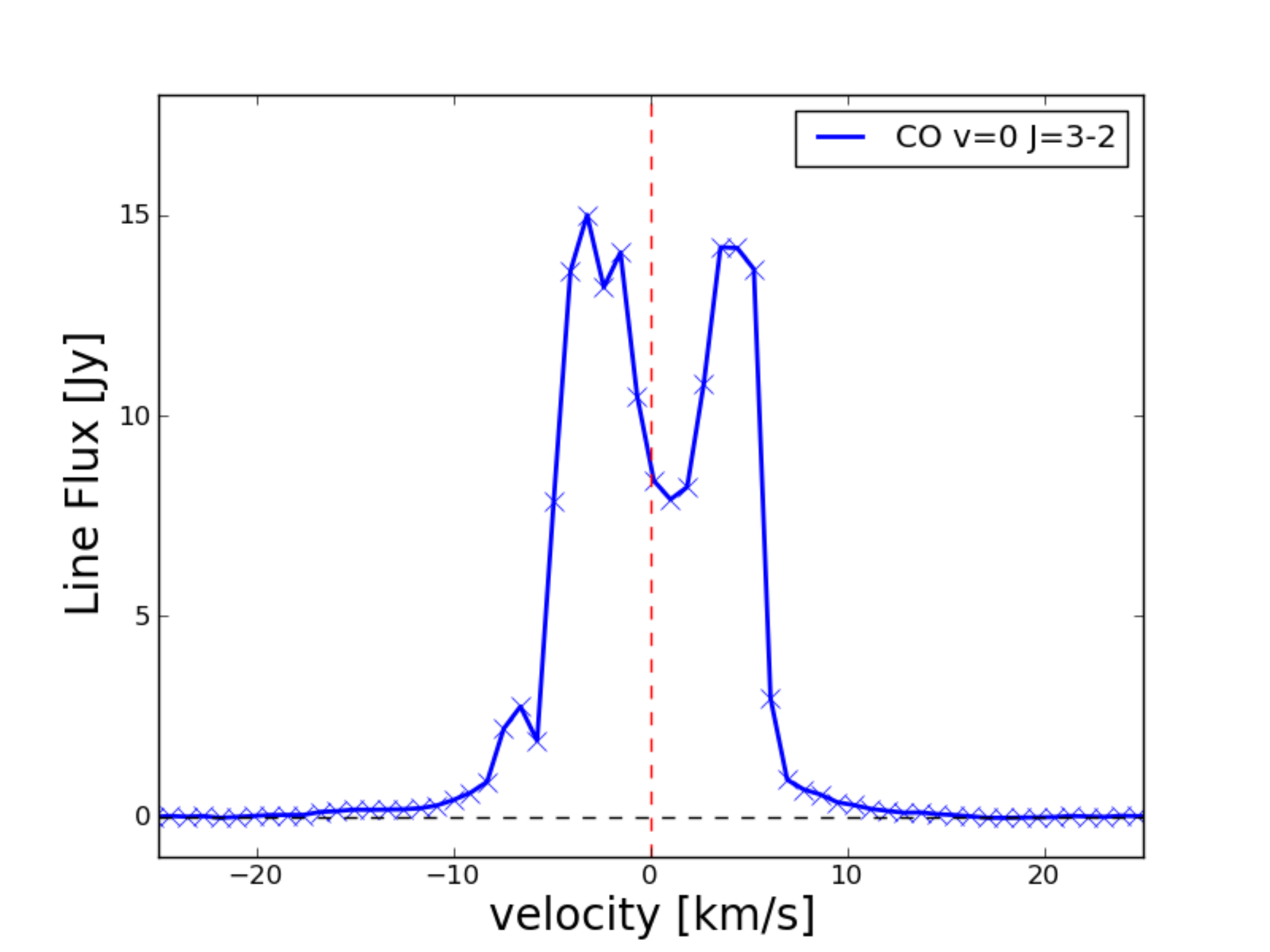}
        \includegraphics[width=9cm]{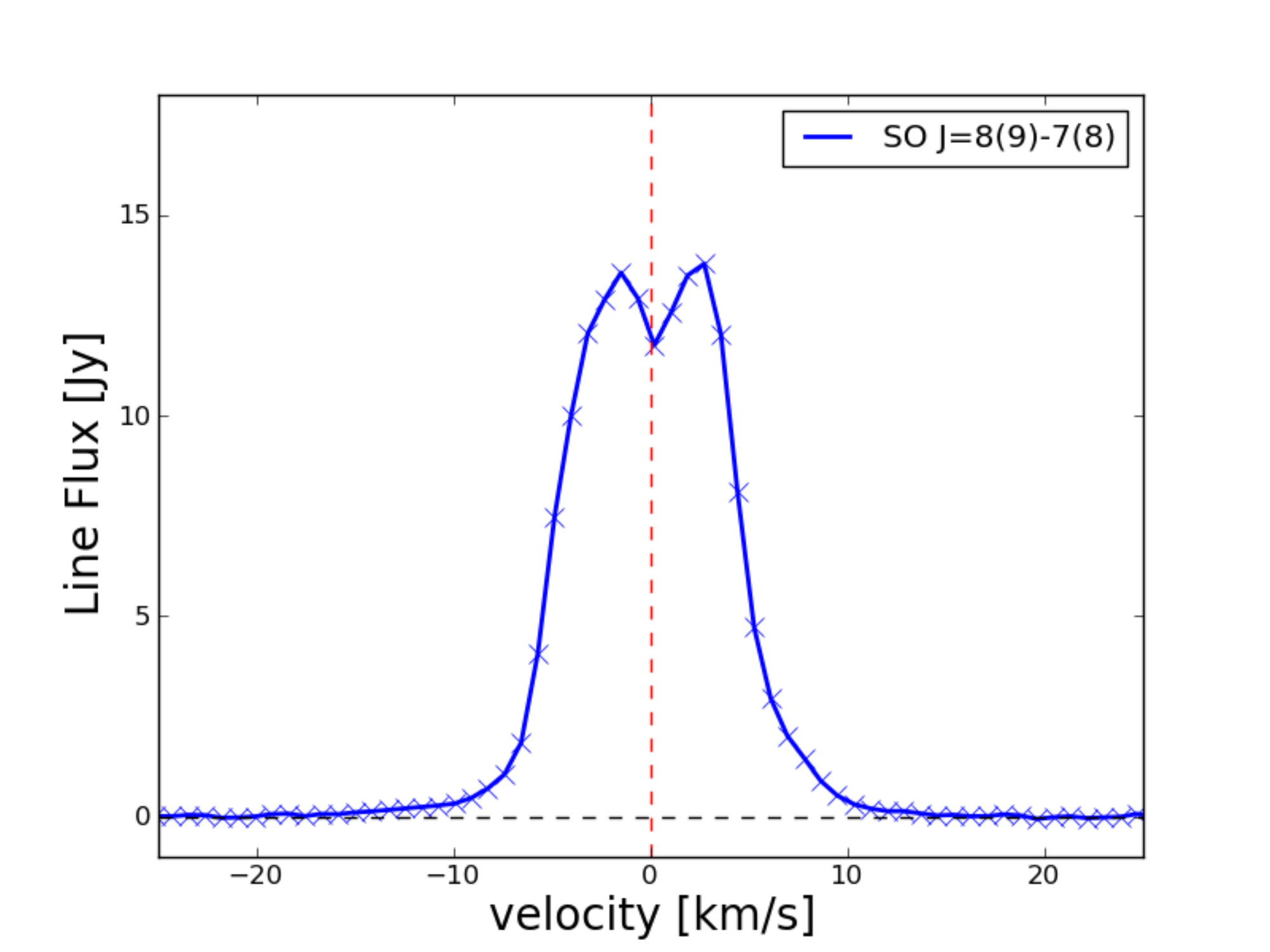}
        \includegraphics[width=9cm]{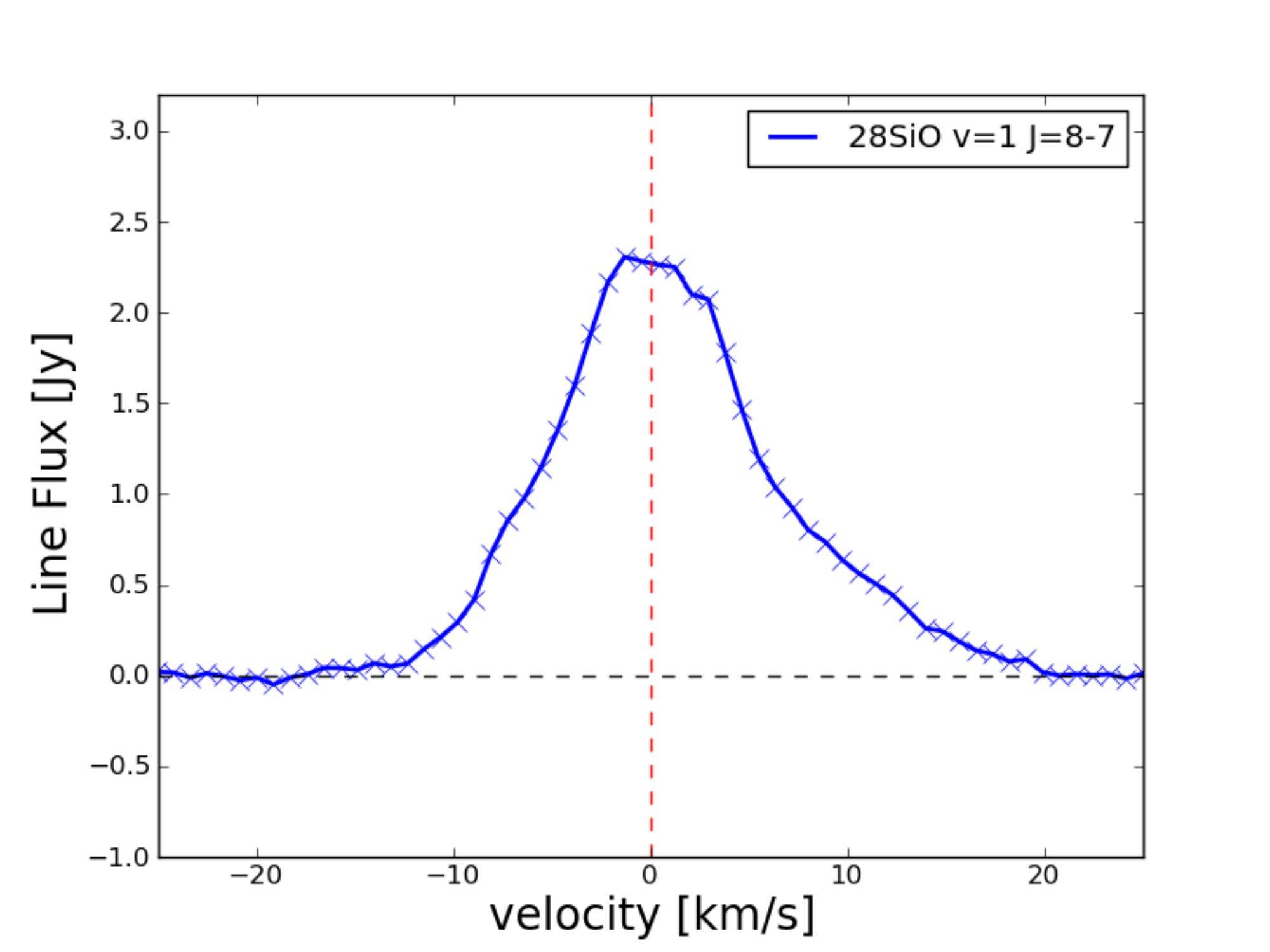}
        \caption{Spectral lines of the considered molecular transitions, for a circular aperture with a radius of 1''. The black dashed line indicates the zero flux level, the red dashed line the line centre (centered at zero velocity with respect to the velocity of the stellar source). Broad line wings are seen in all four lines, but are particularly apparent for the weaker emission line. Such broad line wings are indicative of the potential presence of a differentially rotating disk.
        \label{lines}} 
\end{figure*}

\end{appendix}

\end{document}